\documentclass[11pt]{article}
\usepackage[T1]{fontenc}
\usepackage[utf8]{inputenc}
\usepackage[margin=1in]{geometry}
\usepackage{array,booktabs,tabularx,longtable}
\usepackage{microtype}
\usepackage[table]{xcolor}
\usepackage{tikz}
\usepackage{float}
\usepackage{xurl,seqsplit}
\usepackage{hyperref}
\hypersetup{colorlinks=true,linkcolor=blue!50!black,citecolor=blue!50!black,urlcolor=blue!50!black,
  pdftitle={From JPEG to CUDA Batch: Measuring RGB Input Pipeline Throughput},
  pdfauthor={Vladimir Iglovikov}}
\newcolumntype{Y}{>{\raggedright\arraybackslash}X}
\newcolumntype{P}[1]{>{\raggedright\arraybackslash}p{#1}}
\makeatletter
\let\inputrows\@@input
\makeatother
\newcommand{\repo}{\href{https://github.com/albumentations-team/benchmark/tree/154baa0da1880013941daad1b6bd35c58097f969}{benchmark repository}}
\newcommand{\ProductionCellCount}{1314}
\newcommand{\ProductionPairCount}{438}
\newcommand{\RecipeCount}{73}
\newcommand{\CommonRecipeCount}{34}
\newcommand{\CommonMeanDaliRatio}{1.18}
\newcommand{\CommonDALIFastest}{34}
\newcommand{\CommonMemoryAlbumentationsX}{404}
\newcommand{\CommonBatchRateAlbumentationsX}{17.1}
\newcommand{\CommonBaselineAlbumentationsX}{216}
\newcommand{\CommonMemoryDALIGPU}{1,872}
\newcommand{\CommonBatchRateDALIGPU}{19.7}
\newcommand{\CommonBaselineDALIGPU}{1,872}
\newcommand{\CropAXBatchRate}{17.5}
\newcommand{\CropDALIBatchRate}{19.7}
\newcommand{\PlotKorniaShared}{67}
\newcommand{\PlotTorchVisionShared}{44}
\newcommand{\PlotPillowShared}{46}
\newcommand{\PlotDALIShared}{51}

\newcommand{\PairDALIFaster}{51}
\newcommand{\PairDALISeedCount}{153}
\newcommand{\PairDALISeedFaster}{151}

\newcommand{\MemoryPollMs}{50}

\newcommand{\AllSIMDMedianRatio}{1.16}
\newcommand{\AllSIMDMedianBlurRatio}{16.16}

\newcommand{\FullRunID}{1e7c6db29a2cc0da95eee8f9fce0806f5eb4c5e0c541e1defc784f2f56864d68}
\newcommand{\FullSourceRevision}{fca0c616b4932b4ca50e61b615117e142348f5cb}
\newcommand{\AXRunID}{175d077e5ee85e9785067111573615b3fa7c6bf80641dabf4f5dd72f3a5dae81}
\newcommand{\AXSourceRevision}{0ef0eb42801d246c6f53486c163d6f508452eccf}

\newcommand{\CoverageAlbumentationsX}{125}

\newcommand{\CoverageCatalogSHA}{f617d06554daf23bfbdababe4268ab3393d6214313a3ed4d44ade214be218b9b}

\definecolor{rateabove}{HTML}{236E69}
\definecolor{ratereference}{HTML}{747B82}
\definecolor{ratebelow}{HTML}{ABBAC5}
\definecolor{memmark}{HTML}{454F56}

\newcommand{\ComparisonBar}[6]{%
  \pgfmathsetmacro{\ratepos}{4.12+0.183*#3}%
  \node[anchor=west] at (0,#1) {#2};
  \fill[#5] (4.12,#1-0.15) rectangle (\ratepos,#1+0.15);
  \node[anchor=west,xshift=2pt,fill=white,inner sep=1pt] at (\ratepos,#1)
    {\pgfmathprintnumber[fixed,precision=1,fixed zerofill]{#3}};
  \node[anchor=east,inner xsep=0pt] at (10.68,#1) {#4};
  \node[anchor=west,inner xsep=0pt] at (11.38,#1) {#6};%
}

\title{From JPEG to CUDA Batch:\\Measuring RGB Input Pipeline Throughput}
\author{Vladimir Iglovikov\\Albumentations LLC\\\texttt{vladimir@albumentations.ai}}
\date{November 2026}

\begin{document}
\maketitle

\begin{abstract}
We measure how many ready batches per second eight image input pipelines
deliver to an NVIDIA L4. Each pipeline starts with RGB JPEG files and delivers
synchronized CUDA float16 batches of 256 images. The benchmark contains
\ProductionCellCount{} measurements across \RecipeCount{} recipes. On the
\CommonRecipeCount{} recipes shared by all eight paths, DALI delivered a
median of \CommonBatchRateDALIGPU{} batches/s and AlbumentationsX (AX) delivered
\CommonBatchRateAlbumentationsX{}. DALI was fastest on \CommonDALIFastest{} of
those recipes. Median peak process GPU memory was \CommonMemoryDALIGPU{} MiB
for DALI and \CommonMemoryAlbumentationsX{} MiB for AX.
\end{abstract}

\section{Keep the next batch ready}

The loader can prepare the next batch while the GPU trains on the current
one. If the GPU finishes its step before the next batch is ready, it waits
(Figure~\ref{fig:batch-wait}) \cite{datastalls}.

\begin{figure}[H]
\centering
\begin{tikzpicture}[
  x=1.38cm,y=0.72cm,
  row label/.style={anchor=east,inner xsep=0pt,font=\small},
  heading/.style={anchor=west,font=\small\bfseries,inner xsep=0pt},
  outcome/.style={anchor=east,font=\small\bfseries,inner xsep=0pt},
  prepare/.style={fill=black!10},
  compute/.style={fill=blue!54!black},
  waiting/.style={fill=orange!28,draw=black!55,densely dashed},
]
\draw[black!48,densely dotted] (9,-0.27) -- (9,4.57);

\node[heading] at (0,5.00) {Batches ready in time};
\node[outcome] at (9,5.00) {3 steps completed};
\node[row label] at (-0.20,4.13) {Batch prep};
\node[row label] at (-0.20,3.13) {GPU};

\foreach \start/\finish/\batch in {0/2.4/2,2.4/4.8/3}{
  \path[prepare] (\start,3.84) rectangle (\finish,4.42);
  \node[font=\small] at ({(\start+\finish)/2},4.13) {Prepare batch \batch};
}
\draw[white] (2.4,3.84) -- (2.4,4.42);
\foreach \start/\finish/\batch in {0/3/1,3/6/2,6/9/3}{
  \path[compute] (\start,2.84) rectangle (\finish,3.42);
  \node[font=\small,white] at ({(\start+\finish)/2},3.13) {Train on batch \batch};
}
\foreach \division in {3,6}{
  \draw[white] (\division,2.84) -- (\division,3.42);
}

\node[heading] at (0,2.35) {Batches arrive late};
\node[outcome] at (9,2.35) {2 steps completed};
\node[row label] at (-0.20,1.55) {Batch prep};
\node[row label] at (-0.20,0.38) {GPU};

\foreach \start/\finish/\batch in {0/4.5/2,4.5/9/3}{
  \path[prepare] (\start,1.26) rectangle (\finish,1.84);
  \node[font=\small] at ({(\start+\finish)/2},1.55) {Prepare batch \batch};
}
\draw[white] (4.5,1.26) -- (4.5,1.84);
\draw[->,black!70] (4.5,1.25) -- (4.5,0.68);
\node[anchor=west,font=\scriptsize,inner xsep=3pt] at (4.5,0.97) {Batch 2 ready};
\foreach \start/\finish/\batch in {0/3/1,4.5/7.5/2}{
  \path[compute] (\start,0.09) rectangle (\finish,0.67);
  \node[font=\small,white] at ({(\start+\finish)/2},0.38) {Train on batch \batch};
}
\foreach \start/\finish in {3/4.5,7.5/9}{
  \path[waiting] (\start,0.09) rectangle (\finish,0.67);
  \node[font=\small] at ({(\start+\finish)/2},0.38) {GPU idle};
}

\draw[->,black!55] (0,-0.27) -- (9,-0.27)
  node[anchor=north east,font=\small,inner xsep=0pt] {Time};
\end{tikzpicture}
\caption{Waiting for input reduces training steps per unit time. Both
schedules start with batch 1 ready and use the same GPU step duration.
Below, step 1 ends before batch 2 arrives, so the GPU sits idle until
the loader supplies it. The timings are illustrative.}
\label{fig:batch-wait}
\end{figure}

Input stalls can come from storage, decoding, augmentation, batching, or
transfer \cite{datastalls}. This benchmark measures those stages together,
up to a batch ready for the model. Operator benchmarks time selected routines
\cite{pytorchbenchmark}; training benchmarks such as MLPerf also include
the model and measure progress toward a quality target \cite{mlperf}.

\section{Benchmark design}
\label{sec:workload}

\subsection{Input and timing}

Each measurement covers one implementation, one recipe, and one seed.
After one warm-up batch, the runner times delivery of 32 batches through
the final CUDA synchronization. Batch rate is 32 divided by elapsed
wall-clock seconds. The runner reads the selected ImageNet JPEGs
\cite{imagenet} before timing to reduce the influence of storage. Every
path uses the same file order for a given seed. Decoding remains part of
batch preparation.

\begin{table}[H]
\centering\small
\rowcolors{2}{gray!5}{white}
\begin{tabularx}{\textwidth}{p{3.1cm}Y}
\textbf{Control} & \textbf{Frozen setting} \\ \midrule
Input and access & 10,000 sorted ImageNet-validation JPEGs; prewarmed; shared per-seed permutation \\
Output & CUDA float16 BCHW, \(256\times3\times224\times224\); GPU normalization \\
DataLoader paths & 15 workers; persistent workers; pinned memory; prefetch factor 2 \\
DALI & 15 pipeline threads; native prefetch queue depth 2; native file reader and mixed decoder \\
Timing and repeats & One warm-up batch; 32 measured batches; seeds 137, 138, 139 \\
Machine & Standard \texttt{g2-standard-16}, one NVIDIA L4; 16 vCPUs, 64 GB host memory \\
\end{tabularx}
\caption{Benchmark configuration. Google Cloud specifies Cascade Lake
for G2~\cite{gcp}. Appendix~\ref{app:reproduce} gives the software versions
and source revisions.}
\label{tab:setup}
\end{table}

In the same pass, we sample the GPU memory occupied by the input process
from before pipeline construction through cleanup. The first reading
is the baseline. The peak can include state left by preflight or earlier
recipes. AX uses one native thread per CPU operation; the other paths
use their runtime defaults. Appendix~\ref{app:setup} gives the timing and
randomness details.

\subsection{Execution paths}
\label{sec:paths}

Both Pillow paths use the same adapter with different \texttt{PIL}
packages. Pillow-AllSIMD recorded AVX2 and fast mode, which can change
arithmetic precision~\cite{allsimd}. AX's batch \texttt{InputNormalizer}
uses FP32 arithmetic and returns FP16.

\begin{table}[H]
\centering\small
\rowcolors{2}{gray!5}{white}
\begin{tabularx}{\textwidth}{P{3.4cm}P{2.6cm}YY}
\textbf{Path} & \textbf{JPEG decoder} & \textbf{Before transfer} & \textbf{On GPU} \\ \midrule
AX CPU & SimpleJPEG & Augment; transfer uint8 & Batch InputNormalizer \\
Pillow CPU & Pillow & Augment; transfer uint8 & Normalize \\
Pillow-AllSIMD (fast) & AllSIMD & Augment; transfer uint8 & Normalize \\
TorchVision CPU & TorchVision I/O & Augment; transfer uint8 & Normalize \\
TorchVision GPU & TorchVision I/O & Form batch shape; transfer uint8 & Cast to FP16; augment per image; normalize \\
Kornia CPU & TorchVision I/O & Augment in FP32; transfer FP16 & Normalize \\
Kornia GPU & TorchVision I/O & Form batch shape in FP32; transfer FP16 & Augment batch; normalize \\
DALI GPU & Mixed CPU/GPU & Native reader; DALI manages transfer & Augment and normalize in native graph \\
\end{tabularx}
\caption{Execution paths~\cite{albumentationsx,pillow,torchvision,kornia,dali}.
Throughput includes each path's decoder and its placement of operations.
Each path returns a normalized CUDA float16 batch.}
\label{tab:paths}
\end{table}

\subsection{Which recipes can be compared?}

A recipe specifies the image operations, for example crop followed by
equalization. We align the intended effect, output shape, and parameters
where the APIs permit. Matching a recipe does not guarantee identical
pixels or identical work. Crop offsets, padding, interpolation, and
arithmetic can differ. Appendix~\ref{app:recipes} gives the parameters
and Appendix~\ref{app:semantics} documents the differences.

We use AX measurements from a separate run because an OS package update
overlapped AX's measurements in the full run. Workload, machine settings,
and package locks match. Appendix~\ref{app:reproduce} identifies both
runs. Results retain their source run and cell IDs.

\section{Throughput on the same recipes}

Figure~\ref{fig:delivery} uses the \CommonRecipeCount{} recipes measured by
all eight paths. The full matrix contains \ProductionPairCount{}
implementation--recipe pairs. Pairwise comparisons use the recipes measured
by both paths.

\begin{figure}[H]
\centering
\input{generated/delivery-rows.tex}
\newcommand{\DeliveryRow}[5]{%
  \pgfmathsetmacro{\ratepos}{4.0+0.218*#3}%
  \pgfmathsetmacro{\mempos}{10.1+0.00205*#4}%
  \node[anchor=west] at (0,#1) {#2};
  \fill[#5] (4.0,#1-0.15) rectangle (\ratepos,#1+0.15);
  \node[anchor=west,xshift=2pt,font=\small,fill=white,inner sep=1pt] at (\ratepos,#1)
    {\pgfmathprintnumber[fixed,precision=1,fixed zerofill]{#3}};
  \fill[memmark] (\mempos,#1) circle[radius=2.15pt];
  \node[anchor=west,xshift=4pt,font=\small] at (\mempos,#1)
    {\pgfmathprintnumber[fixed,precision=0]{#4}};%
}
\begin{tikzpicture}[x=1cm,y=0.64cm]
\node[anchor=west,font=\small\bfseries] at (4.0,\DeliveryTopRow+1.1) {Ready batches/s};
\node[anchor=west,font=\small\bfseries] at (10.1,\DeliveryTopRow+1.1) {Peak process GPU memory (MiB)};
\foreach \tick in {0,5,10,15,20}{
  \pgfmathsetmacro{\pos}{4.0+0.218*\tick}
  \draw[black!12] (\pos,-0.48) -- (\pos,\DeliveryTopRow+0.52);
  \node[anchor=north,font=\footnotesize,text=black!75] at (\pos,-0.63) {\tick};
}
\foreach \tick in {0,500,1000,1500,2000}{
  \pgfmathsetmacro{\pos}{10.1+0.00205*\tick}
  \draw[black!12] (\pos,-0.48) -- (\pos,\DeliveryTopRow+0.52);
  \node[anchor=north,font=\footnotesize,text=black!75] at (\pos,-0.63) {\tick};
}
\pgfmathsetmacro{\axratepos}{4.0+0.218*\DeliveryAXRate}
\draw[black!38,densely dashed] (\axratepos,-0.48) -- (\axratepos,\DeliveryTopRow+0.52);
\DeliveryRows
\end{tikzpicture}
\caption{Batch rate and peak process GPU memory on \CommonRecipeCount{} shared
recipes. Each metric takes the median across seeds, then across recipes.
The dashed line marks AX. Teal shows higher rates and pale bars show lower rates.}
\label{fig:delivery}
\end{figure}

The mean of the per-recipe DALI/AX batch-rate ratios was
\CommonMeanDaliRatio\(\times\) on the common set. DALI's median baseline
was \CommonBaselineDALIGPU{} MiB,
equal to its median peak. AX's median baseline was
\CommonBaselineAlbumentationsX{} MiB.

\section{AlbumentationsX against each library}
\label{sec:library-comparisons}

Each comparison uses the recipes measured by every path shown in that
figure. These sets extend the 34 recipes shared by all eight paths.
The bars show median batch rate, with median peak process GPU memory
alongside. The last column says who was faster on more recipes and gives
the count. For each recipe, the comparison
uses the median of three seeds. Appendix~\ref{app:results} lists the
exact recipe sets and results.

\subsection{AX vs DALI}

DALI was faster than AX on \PairDALIFaster{} of \PlotDALIShared{} shared
recipes at the median across seeds. It was also faster in
\PairDALISeedFaster{} of \PairDALISeedCount{} comparisons at the same
recipe and file-order seed. These are repeats on the benchmark workload.
The shared set includes resize, crop, GaussianBlur, Equalize, and Shear.
Elastic and Dithering have no measured DALI path. Table~\ref{tab:paths}
shows how the two paths differ in decoding and placement of operations.

\begin{figure}[H]
\centering
  \begingroup
  \input{generated/comparison-dali.tex}
  \begin{tikzpicture}[x=1cm,y=0.68cm,font=\small]
  \fill[black!4] (8.98,-0.44) rectangle (10.92,\ComparisonTopRow+1.37);
  \foreach \boundary in {8.79,11.11} {
    \draw[black!23] (\boundary,-0.44) -- (\boundary,\ComparisonTopRow+1.37);
  }
  \node[anchor=west,inner xsep=0pt,font=\small\bfseries] at (4.12,\ComparisonTopRow+1.08) {Batches/s};
  \node[anchor=center,inner sep=0pt,font=\small\bfseries] at (9.95,\ComparisonTopRow+1.08) {VRAM};
  \node[anchor=center,inner sep=0pt,font=\footnotesize] at (9.95,\ComparisonTopRow+0.57) {(peak, MiB)};
  \node[anchor=west,inner xsep=0pt,font=\small\bfseries] at (11.38,\ComparisonTopRow+1.08) {Per-recipe result};
  \foreach \tick in {0,5,10,15,20} {
    \pgfmathsetmacro{\pos}{4.12+0.183*\tick}
    \draw[black!12] (\pos,-0.43) -- (\pos,\ComparisonTopRow+0.49);
    \node[anchor=north,font=\footnotesize,text=black!75] at (\pos,-0.54) {\tick};
  }
  \ComparisonRows
  \end{tikzpicture}
  \endgroup

\caption{DALI was faster on all \PlotDALIShared{} recipes measured by both DALI and AX.}
\end{figure}

\subsection{AX vs Pillow}

AX was faster than either Pillow path on most shared recipes. Pillow and
Pillow-AllSIMD run the same CPU adapter, so we compare both against AX
on the \PlotPillowShared{} recipes implemented by all three
paths. The set includes resizing, crops, flips, color adjustments, and
filters such as MedianBlur. Elastic and GaussianNoise have no measured
Pillow mapping in this benchmark.

Pillow-AllSIMD delivered \AllSIMDMedianRatio\(\times\) Pillow's batch rate
at the median across these recipes and \AllSIMDMedianBlurRatio\(\times\) on
MedianBlur (R27). Dithering (R65) is an exception to AX's usual lead here.
AX, Pillow, and Pillow-AllSIMD delivered 3.4, 9.5, and 14.0 batches/s,
respectively. AX diffuses error in floating point, while both Pillow
paths use integer error terms. The resulting pixels can differ
(Appendix~\ref{app:semantics}).

\begin{figure}[H]
\centering
  \begingroup
  \input{generated/comparison-pillow.tex}
  \begin{tikzpicture}[x=1cm,y=0.68cm,font=\small]
  \fill[black!4] (8.98,-0.44) rectangle (10.92,\ComparisonTopRow+1.37);
  \foreach \boundary in {8.79,11.11} {
    \draw[black!23] (\boundary,-0.44) -- (\boundary,\ComparisonTopRow+1.37);
  }
  \node[anchor=west,inner xsep=0pt,font=\small\bfseries] at (4.12,\ComparisonTopRow+1.08) {Batches/s};
  \node[anchor=center,inner sep=0pt,font=\small\bfseries] at (9.95,\ComparisonTopRow+1.08) {VRAM};
  \node[anchor=center,inner sep=0pt,font=\footnotesize] at (9.95,\ComparisonTopRow+0.57) {(peak, MiB)};
  \node[anchor=west,inner xsep=0pt,font=\small\bfseries] at (11.38,\ComparisonTopRow+1.08) {Per-recipe result};
  \foreach \tick in {0,5,10,15,20} {
    \pgfmathsetmacro{\pos}{4.12+0.183*\tick}
    \draw[black!12] (\pos,-0.43) -- (\pos,\ComparisonTopRow+0.49);
    \node[anchor=north,font=\footnotesize,text=black!75] at (\pos,-0.54) {\tick};
  }
  \ComparisonRows
  \end{tikzpicture}
  \endgroup

\caption{AX led both Pillow paths on most of the \PlotPillowShared{} shared recipes.}
\end{figure}

\newpage
\subsection{AX vs Kornia}

AX beat Kornia CPU on every shared recipe. The \PlotKorniaShared{} common recipes include simple crops and flips as
well as Elastic, ThinPlateSpline, illumination changes, and morphology.
Dithering and PixelSpread are among the AX recipes without a measured
Kornia path. The two Kornia paths use the same JPEG decoder. The GPU
path forms a collatable shape on the CPU, then runs the remaining image
operations after transfer (Table~\ref{tab:paths}).

For Elastic (R10), AX and Kornia deliver 11.8 batches/s and 3.2 batches/s
on CPU, respectively. Kornia GPU delivers 9.1 batches/s on that recipe.
The Elastic settings were calibrated by displacement magnitude and
spatial scale, and the implementations still produce different fields.

\begin{figure}[H]
\centering
  \begingroup
  \input{generated/comparison-kornia.tex}
  \begin{tikzpicture}[x=1cm,y=0.68cm,font=\small]
  \fill[black!4] (8.98,-0.44) rectangle (10.92,\ComparisonTopRow+1.37);
  \foreach \boundary in {8.79,11.11} {
    \draw[black!23] (\boundary,-0.44) -- (\boundary,\ComparisonTopRow+1.37);
  }
  \node[anchor=west,inner xsep=0pt,font=\small\bfseries] at (4.12,\ComparisonTopRow+1.08) {Batches/s};
  \node[anchor=center,inner sep=0pt,font=\small\bfseries] at (9.95,\ComparisonTopRow+1.08) {VRAM};
  \node[anchor=center,inner sep=0pt,font=\footnotesize] at (9.95,\ComparisonTopRow+0.57) {(peak, MiB)};
  \node[anchor=west,inner xsep=0pt,font=\small\bfseries] at (11.38,\ComparisonTopRow+1.08) {Per-recipe result};
  \foreach \tick in {0,5,10,15,20} {
    \pgfmathsetmacro{\pos}{4.12+0.183*\tick}
    \draw[black!12] (\pos,-0.43) -- (\pos,\ComparisonTopRow+0.49);
    \node[anchor=north,font=\footnotesize,text=black!75] at (\pos,-0.54) {\tick};
  }
  \ComparisonRows
  \end{tikzpicture}
  \endgroup

\caption{AX was faster than Kornia on nearly all \PlotKorniaShared{} shared recipes.}
\end{figure}

\subsection{AX vs TorchVision}

AX beat both TorchVision paths on most shared recipes. The
\PlotTorchVisionShared{} common recipes include geometry, color
operations, GaussianBlur, Equalize, and JPEG compression. MedianBlur and
CLAHE have no measured TorchVision mapping in this benchmark. The CPU
and GPU paths both use TorchVision I/O, and the GPU path runs enough
geometry on the CPU to form a batch before transfer.

On GaussianNoise (R16), TorchVision went from 4.8 batches/s on CPU to
17.2 on GPU, compared with 13.1 for AX. On Equalize (R22), TorchVision
went from 11.7 batches/s on CPU to 6.5 on GPU, with AX at 15.7. These
are full input pipeline rates. On Elastic (R10), TorchVision GPU delivered
about 0.09 batches/s on all three seeds, versus an AX median of 11.8.
The Elastic mappings align two displacement statistics and can still
produce different fields (Appendix~\ref{app:semantics}).

\begin{figure}[H]
\centering
  \begingroup
  \input{generated/comparison-torchvision.tex}
  \begin{tikzpicture}[x=1cm,y=0.68cm,font=\small]
  \fill[black!4] (8.98,-0.44) rectangle (10.92,\ComparisonTopRow+1.37);
  \foreach \boundary in {8.79,11.11} {
    \draw[black!23] (\boundary,-0.44) -- (\boundary,\ComparisonTopRow+1.37);
  }
  \node[anchor=west,inner xsep=0pt,font=\small\bfseries] at (4.12,\ComparisonTopRow+1.08) {Batches/s};
  \node[anchor=center,inner sep=0pt,font=\small\bfseries] at (9.95,\ComparisonTopRow+1.08) {VRAM};
  \node[anchor=center,inner sep=0pt,font=\footnotesize] at (9.95,\ComparisonTopRow+0.57) {(peak, MiB)};
  \node[anchor=west,inner xsep=0pt,font=\small\bfseries] at (11.38,\ComparisonTopRow+1.08) {Per-recipe result};
  \foreach \tick in {0,5,10,15,20} {
    \pgfmathsetmacro{\pos}{4.12+0.183*\tick}
    \draw[black!12] (\pos,-0.43) -- (\pos,\ComparisonTopRow+0.49);
    \node[anchor=north,font=\footnotesize,text=black!75] at (\pos,-0.54) {\tick};
  }
  \ComparisonRows
  \end{tikzpicture}
  \endgroup

\caption{AX led both TorchVision paths on most of the \PlotTorchVisionShared{} shared recipes.}
\end{figure}

\section{Use the results in a training job}

Start with how often the model needs a batch. For example, a 55\,ms
model step needs about 18.2 batches/s at the same batch size. On the
crop-only recipe (R02), DALI delivered \CropDALIBatchRate{} batches/s and
AX delivered \CropAXBatchRate{}. That makes DALI a candidate to test for
this example. The benchmark measures each loader in isolation. With the
model running, GPU preprocessing can compete for compute and memory, so
measure training throughput again with the candidate pipeline.

If the model still waits for input, locate the slow stage. For storage,
try faster media, larger reads from shards, indexed region reads for
large images or volumes, or caching compressed files. Changing codecs
changes both read volume and decode work. When the input queue empties
and CPU and I/O have spare capacity, try more workers. Deeper prefetch
can absorb short stalls at the cost of larger buffers \cite{pytorchdata}.
If batch preparation still limits delivery, compare decoders and
augmentation paths on the job's actual recipes.

If GPU memory is tight, a smaller device batch can reduce per-step
memory. Gradient accumulation can preserve examples per optimizer update
\cite{pytorchamp}. Batch normalization still uses the samples in each
forward pass. Retune workers and prefetch at the new batch size.
DataLoader prefetch generally uses host memory, and native GPU pipelines
can also buffer on the device. Compare training examples per second
after changing batch size.

\paragraph{Author disclosure.}
The author maintains AlbumentationsX.

\begingroup\footnotesize
\bibliographystyle{plain}
\bibliography{references}
\endgroup

\clearpage
\appendix

\section{Timing, sampling, and output details}
\label{app:setup}

The runner selects the first 10,000 validation JPEGs in lexicographic order
and reads them once before timing. Each seed selects the first 8,448 files
from a PCG64 permutation of that list. After synchronizing the warm-up
batch, the runner times the remaining 8,192 images. Pipeline and worker
setup precede timing. Prefetch can prepare part of the workload before the
timer starts. Reads typically hit the Linux page cache after prewarming.

NVML samples process GPU memory at a nominal \MemoryPollMs{}\,ms interval
and takes readings before pipeline construction and after teardown.
Short peaks can fall between samples. The DataLoader generator is
seeded with zero. The cell seed sets file order, the DALI graph seed,
and the main process's Python, NumPy, and PyTorch RNGs. AX Compose has no
explicit seed in the adapter. Augmentation draws can differ between
implementations.

Normalization uses RGB means \((0.485,0.456,0.406)\) and standard deviations
\((0.229,0.224,0.225)\) on the unit intensity scale. The runner validates
CUDA placement, FP16 type, and BCHW output. The catalog's \texttt{ToTensor}
stage describes this output contract.

The TorchVision and Kornia GPU paths run the prefix through the first
shape-forming resize, crop, or letterbox on the CPU so the images can be stacked.
For a recipe that ends there, only normalization and output preparation run
on the GPU.

The TorchVision adapter typically calls GPU transforms per image and stacks
the results. Its JPEG recipe uses batched CUDA encode and decode, with JPEG
bytes passed through host memory between the calls. Kornia calls its GPU
module on the batch, with \texttt{same\_on\_batch=False} where supported.

\clearpage
\section{Declared recipes and parameters}
\label{app:recipes}

The table lists the image operations and declared sampling. Every recipe
uses GPU normalization and the same tensor output contract
(Section~\ref{sec:workload}). The CSV preserves the full catalog IDs and
stage definitions.

\begingroup\small
\rowcolors{2}{gray!5}{white}
\begin{longtable}{P{4.1cm}P{11.6cm}}
\textbf{Recipe prefix} & \textbf{Catalog parameters and declared sampling} \\ \midrule\endhead
\inputrows generated/recipes.tex
\bottomrule
\end{longtable}
\endgroup

\subsection{Implementation differences}
\label{app:semantics}

\paragraph{Crop and resize.}
R01 and R03 request bilinear resampling without antialiasing. AX uses
\texttt{INTER\_LINEAR}. TorchVision and DALI explicitly disable antialiasing
in those rows. The Pillow paths use bilinear affine resampling, and Kornia
uses bilinear resize with \texttt{align\_corners=False}. In R03, crop sampling,
fallbacks, and rounding can differ. The Pillow adapter tries up to ten
candidate crops and then resizes the whole image if none fits.

RandomCrop224 takes a window at decoded resolution and pads with zeros when
needed. In DALI the slice has integer offsets, with centered padding on
short axes. R50 and R51 resize by long or short side before
cropping~\cite{daliresize}.

\paragraph{VerticalFlip (R05).}
The DALI adapter sets \texttt{horizontal=0} and \texttt{vertical=1}
for vertical-only reflection~\cite{daliflip}.

\paragraph{Shear (R46).}
In pixel-center coordinates, each path maps \(x\) to
\(x+\tan(10^\circ)(y-111.5)\) and leaves \(y\) unchanged. Sampling is bilinear,
with zero-valued neighbors outside the frame.

\paragraph{Affine (R08).}
AX samples rotation in \([-25^\circ,25^\circ]\) with fixed per-axis
translation and shear. TorchVision also samples rotation but interprets the
shear pair as a range for X shear and draws translation using the configured
fractions of the image size. The Pillow paths use a fixed matrix and the
first shear value. Kornia uses a fixed angle and drops the list-valued shear
parameter. DALI uses only scale and shift.

\paragraph{Brightness and blur.}
In R30, AX, DALI, and the Pillow paths use a fixed brightness factor
of 1.2. TorchVision samples a factor in \([0.8,1.2]\). Kornia adds 0.2 in
unit-range intensity coordinates. In GaussianBlur (R15), Pillow's filter
approximates a Gaussian with box filters~\cite{pillowfilters}. R32 applies a
\(5\times5\) box in every measured path. AX, Kornia, and DALI reflect the
boundary, and the Pillow paths extend the edge pixels.

\paragraph{Omitted recipes and grid masks.}
Mosaic, reference-image histogram matching, and box-aware crops require
different inputs. The DALI grid mask drops squares, while AX GridMask
drops stripes.

\paragraph{Dithering (R65).}
All three paths apply binary Floyd--Steinberg error diffusion to grayscale
pixels. AX accumulates error in floating point and quantizes at half scale.
The Pillow paths accumulate integer error terms and use white when the
corrected byte value exceeds 128. These arithmetic choices can produce
different output pixels.

\paragraph{Elastic (R10).}
The AX spline-grid settings were calibrated against the TorchVision and
Kornia displacement fields on CPU, targeting about 0.74 px RMS displacement
and a 10 px autocorrelation scale. The benchmark retains those catalog
settings. Matching these two statistics leaves differences in the fields
and their boundary effects. The calibration procedure is in
\texttt{docs/rgb\_elastic\_calibration.md}.

\clearpage
\section{Results for every measured implementation--recipe pair}
\label{app:results}

Each row gives the median, minimum, and maximum across the same three cells
for throughput and peak GPU memory. The CSV contains unrounded values and
GPU baselines. Recipe IDs link to Appendix~\ref{app:recipes}.

\paragraph{Recipe sets used for the library comparisons.}
The comparisons in Section~\ref{sec:library-comparisons} use the
following catalog IDs.

\begin{description}
\item[DALI] %
  \begingroup
\def\ComparisonRecipes{R01, R02, R03, R04, R05, R06, R07, R08, R11, R12, R13, R14, R15, R16, R17, R18, R19, R22, R23, R24, R25, R27, R29, R30, R31, R32, R33, R37, R42, R43, R46, R49, R50, R51, R52, R53, R58, R59, R60, R61, R62, R63, R64, R66, R67, R68, R69, R70, R71, R72, R73}
\def\ComparisonTopRow{1}
\def\ComparisonRows{%
\ComparisonBar{0}{AlbumentationsX}{16.97934707}{404}{ratereference}{}
\ComparisonBar{1}{DALI}{19.66611910}{1,872}{rateabove}{DALI faster on 51/51}
}
  \ComparisonRecipes
  \endgroup
.
\item[Pillow] %
  \begingroup
\def\ComparisonRecipes{R01, R02, R03, R04, R05, R06, R07, R08, R12, R13, R14, R15, R17, R18, R19, R21, R22, R24, R25, R27, R30, R31, R32, R33, R43, R46, R50, R51, R52, R53, R55, R56, R57, R58, R59, R60, R61, R62, R63, R64, R65, R68, R69, R70, R71, R73}
\def\ComparisonTopRow{2}
\def\ComparisonRows{%
\ComparisonBar{0}{Pillow CPU}{13.62323135}{330}{ratebelow}{AX faster on 44/46}
\ComparisonBar{1}{Pillow-AllSIMD (fast)}{15.30750110}{330}{ratebelow}{AX faster on 41/46}
\ComparisonBar{2}{AlbumentationsX}{17.03550352}{404}{ratereference}{}
}
  \ComparisonRecipes
  \endgroup
.
\item[Kornia] %
  \begingroup
\def\ComparisonRecipes{R01, R02, R03, R04, R05, R06, R07, R08, R09, R10, R11, R12, R13, R14, R15, R16, R17, R18, R19, R20, R21, R22, R23, R24, R25, R26, R27, R28, R29, R30, R31, R32, R33, R34, R35, R36, R37, R38, R39, R40, R41, R42, R43, R44, R45, R46, R47, R49, R50, R51, R52, R53, R54, R58, R59, R60, R61, R62, R63, R64, R66, R67, R68, R69, R70, R71, R73}
\def\ComparisonTopRow{2}
\def\ComparisonRows{%
\ComparisonBar{0}{Kornia CPU}{6.57135378}{292}{ratebelow}{AX faster on 67/67}
\ComparisonBar{1}{Kornia GPU}{9.37493362}{676}{ratebelow}{AX faster on 66/67}
\ComparisonBar{2}{AlbumentationsX}{16.46625203}{404}{ratereference}{}
}
  \ComparisonRecipes
  \endgroup
.
\item[TorchVision] %
  \begingroup
\def\ComparisonRecipes{R01, R02, R03, R04, R05, R06, R07, R08, R09, R10, R11, R12, R13, R15, R16, R17, R18, R19, R20, R21, R22, R23, R24, R25, R30, R31, R37, R46, R48, R49, R50, R51, R52, R53, R58, R59, R60, R61, R62, R63, R68, R69, R70, R73}
\def\ComparisonTopRow{2}
\def\ComparisonRows{%
\ComparisonBar{0}{TorchVision CPU}{13.27193233}{330}{ratebelow}{AX faster on 42/44}
\ComparisonBar{1}{TorchVision GPU}{14.66493130}{544}{ratebelow}{AX faster on 38/44}
\ComparisonBar{2}{AlbumentationsX}{16.97477323}{404}{ratereference}{}
}
  \ComparisonRecipes
  \endgroup
.
\end{description}

\begingroup\small
\rowcolors{2}{gray!5}{white}
\subsection{AlbumentationsX}
\begin{longtable}{lrrrrrr}
\textbf{Recipe} & \multicolumn{3}{c}{\textbf{Throughput (images/s)}} & \multicolumn{3}{c}{\textbf{GPU memory (MiB)}} \\
 & Median & Min & Max & Median & Min & Max \\ \midrule \endhead
\hyperlink{R01}{R01} & 4,586 & 3,647 & 7,627 & 404 & 404 & 404 \\
\hyperlink{R02}{R02} & 4,476 & 4,423 & 4,658 & 404 & 404 & 404 \\
\hyperlink{R03}{R03} & 4,212 & 3,427 & 4,633 & 404 & 404 & 404 \\
\hyperlink{R04}{R04} & 3,775 & 3,721 & 3,967 & 404 & 404 & 404 \\
\hyperlink{R05}{R05} & 4,364 & 3,912 & 4,417 & 404 & 404 & 404 \\
\hyperlink{R06}{R06} & 4,007 & 3,218 & 4,354 & 404 & 404 & 404 \\
\hyperlink{R07}{R07} & 4,344 & 4,048 & 4,531 & 404 & 404 & 404 \\
\hyperlink{R08}{R08} & 4,053 & 3,772 & 4,062 & 404 & 404 & 404 \\
\hyperlink{R09}{R09} & 3,948 & 3,818 & 3,965 & 404 & 404 & 404 \\
\hyperlink{R10}{R10} & 3,010 & 2,316 & 3,065 & 404 & 404 & 404 \\
\hyperlink{R11}{R11} & 3,564 & 2,821 & 3,744 & 404 & 404 & 404 \\
\hyperlink{R12}{R12} & 4,358 & 3,986 & 4,953 & 404 & 404 & 404 \\
\hyperlink{R13}{R13} & 4,686 & 4,651 & 4,815 & 404 & 404 & 404 \\
\hyperlink{R14}{R14} & 4,651 & 4,495 & 4,724 & 404 & 404 & 404 \\
\hyperlink{R15}{R15} & 4,434 & 4,207 & 4,448 & 404 & 404 & 404 \\
\hyperlink{R16}{R16} & 3,344 & 3,323 & 3,373 & 404 & 404 & 404 \\
\hyperlink{R17}{R17} & 4,956 & 4,897 & 4,995 & 404 & 404 & 404 \\
\hyperlink{R18}{R18} & 4,955 & 4,784 & 5,037 & 404 & 404 & 404 \\
\hyperlink{R19}{R19} & 4,586 & 4,500 & 4,847 & 404 & 404 & 404 \\
\hyperlink{R20}{R20} & 4,404 & 4,241 & 4,460 & 404 & 404 & 404 \\
\hyperlink{R21}{R21} & 4,057 & 3,874 & 4,227 & 404 & 404 & 404 \\
\hyperlink{R22}{R22} & 4,008 & 3,843 & 4,084 & 404 & 404 & 404 \\
\hyperlink{R23}{R23} & 4,722 & 4,705 & 4,903 & 404 & 404 & 404 \\
\hyperlink{R24}{R24} & 4,036 & 3,907 & 4,050 & 404 & 404 & 404 \\
\hyperlink{R25}{R25} & 4,643 & 3,449 & 4,742 & 404 & 404 & 404 \\
\hyperlink{R26}{R26} & 4,420 & 4,329 & 4,523 & 404 & 404 & 404 \\
\hyperlink{R27}{R27} & 3,773 & 3,672 & 3,891 & 404 & 404 & 404 \\
\hyperlink{R28}{R28} & 4,215 & 3,183 & 4,261 & 404 & 404 & 404 \\
\hyperlink{R29}{R29} & 3,161 & 3,063 & 3,253 & 404 & 404 & 404 \\
\hyperlink{R30}{R30} & 4,592 & 4,299 & 4,736 & 404 & 404 & 404 \\
\hyperlink{R31}{R31} & 4,539 & 4,179 & 4,554 & 404 & 404 & 404 \\
\hyperlink{R32}{R32} & 4,650 & 4,622 & 4,714 & 404 & 404 & 404 \\
\hyperlink{R33}{R33} & 4,906 & 3,587 & 4,993 & 404 & 404 & 404 \\
\hyperlink{R34}{R34} & 2,916 & 2,879 & 3,849 & 404 & 404 & 404 \\
\hyperlink{R35}{R35} & 4,044 & 3,882 & 4,091 & 404 & 404 & 404 \\
\hyperlink{R36}{R36} & 3,715 & 2,823 & 3,900 & 404 & 404 & 404 \\
\hyperlink{R37}{R37} & 4,061 & 3,835 & 4,071 & 404 & 404 & 404 \\
\hyperlink{R38}{R38} & 2,591 & 2,438 & 2,660 & 404 & 404 & 404 \\
\hyperlink{R39}{R39} & 2,295 & 2,216 & 2,299 & 404 & 404 & 404 \\
\hyperlink{R40}{R40} & 2,450 & 2,445 & 2,520 & 404 & 404 & 404 \\
\hyperlink{R41}{R41} & 3,926 & 3,763 & 4,136 & 404 & 404 & 404 \\
\hyperlink{R42}{R42} & 4,600 & 4,525 & 4,688 & 404 & 404 & 404 \\
\hyperlink{R43}{R43} & 3,791 & 3,184 & 4,129 & 404 & 404 & 404 \\
\hyperlink{R44}{R44} & 3,825 & 3,733 & 3,839 & 404 & 404 & 404 \\
\hyperlink{R45}{R45} & 3,835 & 3,562 & 3,863 & 404 & 404 & 404 \\
\hyperlink{R46}{R46} & 3,861 & 3,738 & 4,071 & 404 & 404 & 404 \\
\hyperlink{R47}{R47} & 3,050 & 2,356 & 3,153 & 404 & 404 & 404 \\
\hyperlink{R48}{R48} & 3,420 & 3,399 & 3,469 & 404 & 404 & 404 \\
\hyperlink{R49}{R49} & 3,730 & 3,551 & 3,742 & 404 & 404 & 404 \\
\hyperlink{R50}{R50} & 3,768 & 3,512 & 3,796 & 404 & 404 & 404 \\
\hyperlink{R51}{R51} & 3,204 & 3,138 & 3,228 & 404 & 404 & 404 \\
\hyperlink{R52}{R52} & 4,919 & 4,754 & 4,993 & 404 & 404 & 404 \\
\hyperlink{R53}{R53} & 4,506 & 4,480 & 4,621 & 404 & 404 & 404 \\
\hyperlink{R54}{R54} & 4,307 & 3,516 & 4,430 & 404 & 404 & 404 \\
\hyperlink{R55}{R55} & 4,522 & 4,124 & 4,570 & 404 & 404 & 404 \\
\hyperlink{R56}{R56} & 4,402 & 4,367 & 4,502 & 404 & 404 & 404 \\
\hyperlink{R57}{R57} & 4,181 & 4,092 & 4,262 & 404 & 404 & 404 \\
\hyperlink{R58}{R58} & 4,521 & 4,137 & 4,756 & 404 & 404 & 404 \\
\hyperlink{R59}{R59} & 4,625 & 4,360 & 4,628 & 404 & 404 & 404 \\
\hyperlink{R60}{R60} & 4,404 & 4,372 & 4,510 & 404 & 404 & 404 \\
\hyperlink{R61}{R61} & 4,558 & 4,205 & 4,736 & 404 & 404 & 404 \\
\hyperlink{R62}{R62} & 4,235 & 4,200 & 4,395 & 404 & 404 & 404 \\
\hyperlink{R63}{R63} & 4,475 & 4,188 & 4,699 & 404 & 404 & 404 \\
\hyperlink{R64}{R64} & 4,076 & 3,286 & 4,144 & 404 & 404 & 404 \\
\hyperlink{R65}{R65} & 882 & 873 & 905 & 404 & 404 & 404 \\
\hyperlink{R66}{R66} & 3,348 & 3,348 & 4,451 & 404 & 404 & 404 \\
\hyperlink{R67}{R67} & 4,369 & 4,313 & 4,518 & 404 & 404 & 404 \\
\hyperlink{R68}{R68} & 3,563 & 3,561 & 3,636 & 404 & 404 & 404 \\
\hyperlink{R69}{R69} & 4,319 & 3,255 & 4,367 & 404 & 404 & 404 \\
\hyperlink{R70}{R70} & 4,347 & 4,171 & 4,498 & 404 & 404 & 404 \\
\hyperlink{R71}{R71} & 4,242 & 3,973 & 4,247 & 404 & 404 & 404 \\
\hyperlink{R72}{R72} & 3,495 & 3,473 & 3,574 & 404 & 404 & 404 \\
\hyperlink{R73}{R73} & 3,718 & 2,869 & 3,870 & 404 & 404 & 404 \\
\bottomrule\end{longtable}
\subsection{Pillow CPU}
\begin{longtable}{lrrrrrr}
\textbf{Recipe} & \multicolumn{3}{c}{\textbf{Throughput (images/s)}} & \multicolumn{3}{c}{\textbf{GPU memory (MiB)}} \\
 & Median & Min & Max & Median & Min & Max \\ \midrule \endhead
\hyperlink{R01}{R01} & 2,708 & 2,597 & 3,228 & 330 & 330 & 330 \\
\hyperlink{R02}{R02} & 4,341 & 4,282 & 4,346 & 330 & 330 & 330 \\
\hyperlink{R03}{R03} & 2,654 & 2,604 & 2,698 & 330 & 330 & 330 \\
\hyperlink{R04}{R04} & 4,072 & 2,535 & 4,150 & 330 & 330 & 330 \\
\hyperlink{R05}{R05} & 4,229 & 4,204 & 4,284 & 330 & 330 & 330 \\
\hyperlink{R06}{R06} & 3,868 & 3,707 & 4,000 & 330 & 330 & 330 \\
\hyperlink{R07}{R07} & 3,956 & 3,872 & 3,972 & 330 & 330 & 330 \\
\hyperlink{R08}{R08} & 2,870 & 2,864 & 2,907 & 330 & 330 & 330 \\
\hyperlink{R12}{R12} & 4,068 & 2,523 & 4,073 & 330 & 330 & 330 \\
\hyperlink{R13}{R13} & 4,071 & 2,478 & 4,096 & 330 & 330 & 330 \\
\hyperlink{R14}{R14} & 3,736 & 2,492 & 3,763 & 330 & 330 & 330 \\
\hyperlink{R15}{R15} & 2,487 & 2,471 & 2,491 & 330 & 330 & 330 \\
\hyperlink{R17}{R17} & 4,025 & 3,881 & 4,162 & 330 & 330 & 330 \\
\hyperlink{R18}{R18} & 4,019 & 3,839 & 4,126 & 330 & 330 & 330 \\
\hyperlink{R19}{R19} & 3,899 & 3,830 & 3,978 & 330 & 330 & 330 \\
\hyperlink{R21}{R21} & 3,675 & 2,315 & 3,791 & 330 & 330 & 330 \\
\hyperlink{R22}{R22} & 3,737 & 3,683 & 3,781 & 330 & 330 & 330 \\
\hyperlink{R24}{R24} & 3,566 & 3,490 & 3,572 & 330 & 330 & 330 \\
\hyperlink{R25}{R25} & 4,002 & 3,758 & 4,030 & 330 & 330 & 330 \\
\hyperlink{R27}{R27} & 243 & 209 & 245 & 330 & 330 & 330 \\
\hyperlink{R30}{R30} & 3,918 & 3,846 & 3,960 & 330 & 330 & 330 \\
\hyperlink{R31}{R31} & 3,735 & 2,309 & 3,752 & 330 & 330 & 330 \\
\hyperlink{R32}{R32} & 3,409 & 3,343 & 3,504 & 330 & 330 & 330 \\
\hyperlink{R33}{R33} & 4,008 & 4,008 & 4,104 & 330 & 330 & 330 \\
\hyperlink{R43}{R43} & 3,751 & 3,724 & 3,827 & 330 & 330 & 330 \\
\hyperlink{R46}{R46} & 2,663 & 2,659 & 2,694 & 330 & 330 & 330 \\
\hyperlink{R50}{R50} & 1,260 & 1,236 & 1,267 & 330 & 330 & 330 \\
\hyperlink{R51}{R51} & 815 & 811 & 822 & 330 & 330 & 330 \\
\hyperlink{R52}{R52} & 4,179 & 4,178 & 4,233 & 330 & 330 & 330 \\
\hyperlink{R53}{R53} & 4,128 & 4,057 & 4,170 & 330 & 330 & 330 \\
\hyperlink{R55}{R55} & 2,863 & 2,800 & 2,959 & 330 & 330 & 330 \\
\hyperlink{R56}{R56} & 2,849 & 2,811 & 2,896 & 330 & 330 & 330 \\
\hyperlink{R57}{R57} & 2,241 & 2,235 & 2,335 & 330 & 330 & 330 \\
\hyperlink{R58}{R58} & 4,187 & 4,090 & 4,253 & 330 & 330 & 330 \\
\hyperlink{R59}{R59} & 2,334 & 1,585 & 2,344 & 330 & 330 & 330 \\
\hyperlink{R60}{R60} & 2,611 & 1,784 & 2,683 & 330 & 330 & 330 \\
\hyperlink{R61}{R61} & 2,883 & 2,862 & 2,961 & 330 & 330 & 330 \\
\hyperlink{R62}{R62} & 4,217 & 4,180 & 4,285 & 330 & 330 & 330 \\
\hyperlink{R63}{R63} & 2,587 & 2,584 & 2,683 & 330 & 330 & 330 \\
\hyperlink{R64}{R64} & 2,381 & 1,603 & 2,396 & 330 & 330 & 330 \\
\hyperlink{R65}{R65} & 2,440 & 2,409 & 2,591 & 330 & 330 & 330 \\
\hyperlink{R68}{R68} & 1,143 & 1,048 & 1,150 & 330 & 330 & 330 \\
\hyperlink{R69}{R69} & 1,937 & 1,915 & 1,970 & 330 & 330 & 330 \\
\hyperlink{R70}{R70} & 2,662 & 2,596 & 2,758 & 330 & 330 & 330 \\
\hyperlink{R71}{R71} & 1,836 & 1,732 & 2,016 & 330 & 330 & 330 \\
\hyperlink{R73}{R73} & 1,952 & 1,909 & 1,981 & 330 & 330 & 330 \\
\bottomrule\end{longtable}
\subsection{TorchVision CPU}
\begin{longtable}{lrrrrrr}
\textbf{Recipe} & \multicolumn{3}{c}{\textbf{Throughput (images/s)}} & \multicolumn{3}{c}{\textbf{GPU memory (MiB)}} \\
 & Median & Min & Max & Median & Min & Max \\ \midrule \endhead
\hyperlink{R01}{R01} & 3,787 & 3,691 & 5,945 & 330 & 330 & 330 \\
\hyperlink{R02}{R02} & 4,217 & 3,131 & 4,409 & 330 & 330 & 330 \\
\hyperlink{R03}{R03} & 3,490 & 2,647 & 3,607 & 330 & 330 & 330 \\
\hyperlink{R04}{R04} & 4,245 & 4,244 & 4,256 & 330 & 330 & 330 \\
\hyperlink{R05}{R05} & 4,169 & 4,087 & 4,217 & 330 & 330 & 330 \\
\hyperlink{R06}{R06} & 3,563 & 3,470 & 3,565 & 330 & 330 & 330 \\
\hyperlink{R07}{R07} & 2,519 & 2,485 & 2,590 & 330 & 330 & 330 \\
\hyperlink{R08}{R08} & 2,396 & 2,353 & 2,485 & 330 & 330 & 330 \\
\hyperlink{R09}{R09} & 2,237 & 2,143 & 2,260 & 330 & 330 & 330 \\
\hyperlink{R10}{R10} & 231 & 228 & 236 & 330 & 330 & 330 \\
\hyperlink{R11}{R11} & 1,203 & 1,199 & 1,220 & 330 & 330 & 330 \\
\hyperlink{R12}{R12} & 4,030 & 3,849 & 4,122 & 330 & 330 & 330 \\
\hyperlink{R13}{R13} & 3,877 & 3,824 & 3,926 & 330 & 330 & 330 \\
\hyperlink{R15}{R15} & 2,204 & 2,069 & 2,577 & 330 & 330 & 330 \\
\hyperlink{R16}{R16} & 1,236 & 1,218 & 1,267 & 330 & 330 & 330 \\
\hyperlink{R17}{R17} & 4,232 & 4,089 & 4,276 & 330 & 330 & 330 \\
\hyperlink{R18}{R18} & 4,264 & 3,104 & 4,439 & 330 & 330 & 330 \\
\hyperlink{R19}{R19} & 3,391 & 2,838 & 3,645 & 330 & 330 & 330 \\
\hyperlink{R20}{R20} & 1,969 & 1,503 & 2,300 & 330 & 330 & 330 \\
\hyperlink{R21}{R21} & 2,820 & 2,646 & 2,837 & 330 & 330 & 330 \\
\hyperlink{R22}{R22} & 2,985 & 2,918 & 3,104 & 330 & 330 & 330 \\
\hyperlink{R23}{R23} & 3,943 & 3,715 & 4,138 & 330 & 330 & 330 \\
\hyperlink{R24}{R24} & 3,404 & 3,383 & 3,451 & 330 & 330 & 330 \\
\hyperlink{R25}{R25} & 3,151 & 2,341 & 3,173 & 330 & 330 & 330 \\
\hyperlink{R30}{R30} & 3,755 & 3,643 & 3,759 & 330 & 330 & 330 \\
\hyperlink{R31}{R31} & 3,431 & 2,603 & 3,496 & 330 & 330 & 330 \\
\hyperlink{R37}{R37} & 1,377 & 1,104 & 1,416 & 330 & 330 & 330 \\
\hyperlink{R46}{R46} & 2,788 & 2,772 & 2,814 & 330 & 330 & 330 \\
\hyperlink{R48}{R48} & 1,161 & 1,150 & 1,181 & 330 & 330 & 330 \\
\hyperlink{R49}{R49} & 1,220 & 1,206 & 1,238 & 330 & 330 & 330 \\
\hyperlink{R50}{R50} & 3,250 & 2,476 & 3,421 & 330 & 330 & 330 \\
\hyperlink{R51}{R51} & 3,133 & 2,961 & 3,160 & 330 & 330 & 330 \\
\hyperlink{R52}{R52} & 4,101 & 4,027 & 4,199 & 330 & 330 & 330 \\
\hyperlink{R53}{R53} & 4,049 & 3,250 & 4,246 & 330 & 330 & 330 \\
\hyperlink{R58}{R58} & 4,311 & 3,152 & 4,479 & 330 & 330 & 330 \\
\hyperlink{R59}{R59} & 2,916 & 2,767 & 3,768 & 330 & 330 & 330 \\
\hyperlink{R60}{R60} & 3,779 & 3,679 & 3,866 & 330 & 330 & 330 \\
\hyperlink{R61}{R61} & 3,839 & 3,499 & 3,963 & 330 & 330 & 330 \\
\hyperlink{R62}{R62} & 4,267 & 4,126 & 4,475 & 330 & 330 & 330 \\
\hyperlink{R63}{R63} & 3,802 & 3,344 & 3,833 & 330 & 330 & 330 \\
\hyperlink{R68}{R68} & 3,170 & 3,038 & 3,176 & 330 & 330 & 330 \\
\hyperlink{R69}{R69} & 3,309 & 2,547 & 3,314 & 330 & 330 & 330 \\
\hyperlink{R70}{R70} & 3,452 & 2,788 & 3,600 & 330 & 330 & 330 \\
\hyperlink{R73}{R73} & 2,044 & 1,949 & 2,504 & 330 & 330 & 330 \\
\bottomrule\end{longtable}
\subsection{TorchVision GPU}
\begin{longtable}{lrrrrrr}
\textbf{Recipe} & \multicolumn{3}{c}{\textbf{Throughput (images/s)}} & \multicolumn{3}{c}{\textbf{GPU memory (MiB)}} \\
 & Median & Min & Max & Median & Min & Max \\ \midrule \endhead
\hyperlink{R01}{R01} & 3,887 & 3,819 & 5,865 & 386 & 386 & 386 \\
\hyperlink{R02}{R02} & 4,353 & 3,444 & 4,528 & 386 & 386 & 386 \\
\hyperlink{R03}{R03} & 3,691 & 3,663 & 3,819 & 386 & 386 & 386 \\
\hyperlink{R04}{R04} & 4,482 & 4,454 & 4,700 & 544 & 544 & 544 \\
\hyperlink{R05}{R05} & 4,556 & 4,292 & 4,557 & 544 & 544 & 544 \\
\hyperlink{R06}{R06} & 3,756 & 3,755 & 3,890 & 386 & 386 & 386 \\
\hyperlink{R07}{R07} & 1,523 & 1,450 & 1,526 & 564 & 564 & 564 \\
\hyperlink{R08}{R08} & 1,515 & 1,462 & 1,516 & 582 & 582 & 582 \\
\hyperlink{R09}{R09} & 875 & 864 & 904 & 582 & 582 & 582 \\
\hyperlink{R10}{R10} & 22 & 22 & 22 & 562 & 562 & 562 \\
\hyperlink{R11}{R11} & 766 & 765 & 773 & 548 & 548 & 548 \\
\hyperlink{R12}{R12} & 4,550 & 4,448 & 4,564 & 544 & 544 & 544 \\
\hyperlink{R13}{R13} & 4,473 & 4,319 & 4,560 & 484 & 484 & 484 \\
\hyperlink{R15}{R15} & 2,918 & 2,888 & 2,929 & 544 & 544 & 544 \\
\hyperlink{R16}{R16} & 4,403 & 4,373 & 4,663 & 544 & 544 & 544 \\
\hyperlink{R17}{R17} & 4,455 & 4,179 & 4,647 & 544 & 544 & 544 \\
\hyperlink{R18}{R18} & 4,503 & 4,411 & 4,732 & 544 & 544 & 544 \\
\hyperlink{R19}{R19} & 4,432 & 4,400 & 4,526 & 544 & 544 & 544 \\
\hyperlink{R20}{R20} & 3,335 & 3,307 & 3,386 & 544 & 544 & 544 \\
\hyperlink{R21}{R21} & 4,270 & 4,199 & 4,323 & 544 & 544 & 544 \\
\hyperlink{R22}{R22} & 1,671 & 1,632 & 1,690 & 544 & 544 & 544 \\
\hyperlink{R23}{R23} & 3,001 & 2,997 & 3,065 & 544 & 544 & 544 \\
\hyperlink{R24}{R24} & 2,428 & 2,388 & 2,434 & 464 & 464 & 464 \\
\hyperlink{R25}{R25} & 4,455 & 4,343 & 4,520 & 544 & 544 & 544 \\
\hyperlink{R30}{R30} & 4,348 & 3,294 & 4,589 & 544 & 544 & 544 \\
\hyperlink{R31}{R31} & 3,595 & 3,593 & 3,627 & 544 & 544 & 544 \\
\hyperlink{R37}{R37} & 968 & 959 & 968 & 546 & 546 & 546 \\
\hyperlink{R46}{R46} & 2,060 & 1,794 & 2,065 & 566 & 566 & 566 \\
\hyperlink{R48}{R48} & 702 & 701 & 703 & 548 & 548 & 548 \\
\hyperlink{R49}{R49} & 771 & 769 & 773 & 548 & 548 & 548 \\
\hyperlink{R50}{R50} & 3,258 & 2,668 & 3,485 & 386 & 386 & 386 \\
\hyperlink{R51}{R51} & 3,091 & 2,306 & 3,230 & 386 & 386 & 386 \\
\hyperlink{R52}{R52} & 4,439 & 4,434 & 4,554 & 460 & 460 & 460 \\
\hyperlink{R53}{R53} & 4,301 & 3,283 & 4,480 & 544 & 544 & 544 \\
\hyperlink{R58}{R58} & 4,439 & 4,256 & 4,545 & 386 & 386 & 386 \\
\hyperlink{R59}{R59} & 4,019 & 3,822 & 4,061 & 492 & 492 & 492 \\
\hyperlink{R60}{R60} & 4,115 & 3,985 & 4,189 & 544 & 544 & 544 \\
\hyperlink{R61}{R61} & 3,882 & 2,977 & 3,903 & 386 & 386 & 386 \\
\hyperlink{R62}{R62} & 4,374 & 4,342 & 4,388 & 386 & 386 & 386 \\
\hyperlink{R63}{R63} & 4,006 & 3,931 & 4,019 & 544 & 544 & 544 \\
\hyperlink{R68}{R68} & 3,278 & 3,236 & 3,284 & 386 & 386 & 386 \\
\hyperlink{R69}{R69} & 3,753 & 3,645 & 3,793 & 544 & 544 & 544 \\
\hyperlink{R70}{R70} & 3,184 & 3,136 & 3,344 & 544 & 544 & 544 \\
\hyperlink{R73}{R73} & 1,937 & 1,922 & 1,955 & 566 & 566 & 566 \\
\bottomrule\end{longtable}
\subsection{Kornia CPU}
\begin{longtable}{lrrrrrr}
\textbf{Recipe} & \multicolumn{3}{c}{\textbf{Throughput (images/s)}} & \multicolumn{3}{c}{\textbf{GPU memory (MiB)}} \\
 & Median & Min & Max & Median & Min & Max \\ \midrule \endhead
\hyperlink{R01}{R01} & 2,431 & 1,622 & 4,314 & 292 & 292 & 292 \\
\hyperlink{R02}{R02} & 2,431 & 2,320 & 2,497 & 292 & 292 & 292 \\
\hyperlink{R03}{R03} & 2,068 & 1,326 & 2,185 & 292 & 292 & 292 \\
\hyperlink{R04}{R04} & 2,355 & 2,269 & 2,378 & 292 & 292 & 292 \\
\hyperlink{R05}{R05} & 2,354 & 2,239 & 2,396 & 292 & 292 & 292 \\
\hyperlink{R06}{R06} & 1,972 & 1,825 & 2,021 & 292 & 292 & 292 \\
\hyperlink{R07}{R07} & 1,516 & 1,496 & 1,560 & 292 & 292 & 292 \\
\hyperlink{R08}{R08} & 1,527 & 1,096 & 1,600 & 292 & 292 & 292 \\
\hyperlink{R09}{R09} & 1,460 & 1,425 & 1,554 & 292 & 292 & 292 \\
\hyperlink{R10}{R10} & 826 & 818 & 861 & 292 & 292 & 292 \\
\hyperlink{R11}{R11} & 1,013 & 1,013 & 1,054 & 292 & 292 & 292 \\
\hyperlink{R12}{R12} & 2,282 & 2,259 & 2,419 & 292 & 292 & 292 \\
\hyperlink{R13}{R13} & 2,185 & 2,184 & 2,251 & 292 & 292 & 292 \\
\hyperlink{R14}{R14} & 2,090 & 1,490 & 2,146 & 292 & 292 & 292 \\
\hyperlink{R15}{R15} & 1,311 & 1,218 & 1,337 & 292 & 292 & 292 \\
\hyperlink{R16}{R16} & 1,895 & 1,338 & 1,935 & 292 & 292 & 292 \\
\hyperlink{R17}{R17} & 2,343 & 2,314 & 2,354 & 292 & 292 & 292 \\
\hyperlink{R18}{R18} & 1,792 & 1,721 & 1,798 & 292 & 292 & 292 \\
\hyperlink{R19}{R19} & 1,917 & 1,909 & 1,968 & 292 & 292 & 292 \\
\hyperlink{R20}{R20} & 1,309 & 1,264 & 1,353 & 292 & 292 & 292 \\
\hyperlink{R21}{R21} & 1,946 & 1,930 & 2,083 & 292 & 292 & 292 \\
\hyperlink{R22}{R22} & 1,679 & 1,630 & 1,705 & 292 & 292 & 292 \\
\hyperlink{R23}{R23} & 1,962 & 1,871 & 2,071 & 292 & 292 & 292 \\
\hyperlink{R24}{R24} & 680 & 679 & 686 & 292 & 292 & 292 \\
\hyperlink{R25}{R25} & 1,821 & 1,270 & 1,914 & 292 & 292 & 292 \\
\hyperlink{R26}{R26} & 2,070 & 1,400 & 2,087 & 292 & 292 & 292 \\
\hyperlink{R27}{R27} & 402 & 402 & 413 & 292 & 292 & 292 \\
\hyperlink{R28}{R28} & 1,540 & 1,499 & 1,567 & 292 & 292 & 292 \\
\hyperlink{R29}{R29} & 871 & 867 & 888 & 292 & 292 & 292 \\
\hyperlink{R30}{R30} & 2,153 & 1,522 & 2,159 & 292 & 292 & 292 \\
\hyperlink{R31}{R31} & 2,186 & 2,183 & 2,202 & 292 & 292 & 292 \\
\hyperlink{R32}{R32} & 1,420 & 1,334 & 1,422 & 292 & 292 & 292 \\
\hyperlink{R33}{R33} & 2,232 & 1,454 & 2,257 & 292 & 292 & 292 \\
\hyperlink{R34}{R34} & 2,077 & 2,028 & 2,080 & 292 & 292 & 292 \\
\hyperlink{R35}{R35} & 1,669 & 1,628 & 1,701 & 292 & 292 & 292 \\
\hyperlink{R36}{R36} & 1,634 & 1,630 & 1,649 & 292 & 292 & 292 \\
\hyperlink{R37}{R37} & 1,148 & 1,135 & 1,149 & 292 & 292 & 292 \\
\hyperlink{R38}{R38} & 490 & 474 & 494 & 292 & 292 & 292 \\
\hyperlink{R39}{R39} & 488 & 480 & 494 & 292 & 292 & 292 \\
\hyperlink{R40}{R40} & 850 & 850 & 858 & 292 & 292 & 292 \\
\hyperlink{R41}{R41} & 1,653 & 1,645 & 1,658 & 292 & 292 & 292 \\
\hyperlink{R42}{R42} & 1,708 & 1,700 & 1,723 & 292 & 292 & 292 \\
\hyperlink{R43}{R43} & 1,832 & 1,797 & 1,890 & 292 & 292 & 292 \\
\hyperlink{R44}{R44} & 1,121 & 827 & 1,156 & 292 & 292 & 292 \\
\hyperlink{R45}{R45} & 1,405 & 1,358 & 1,423 & 292 & 292 & 292 \\
\hyperlink{R46}{R46} & 1,716 & 1,693 & 1,726 & 292 & 292 & 292 \\
\hyperlink{R47}{R47} & 660 & 657 & 665 & 292 & 292 & 292 \\
\hyperlink{R49}{R49} & 1,011 & 1,001 & 1,012 & 292 & 292 & 292 \\
\hyperlink{R50}{R50} & 1,386 & 1,299 & 1,388 & 292 & 292 & 292 \\
\hyperlink{R51}{R51} & 1,023 & 1,022 & 1,030 & 292 & 292 & 292 \\
\hyperlink{R52}{R52} & 2,398 & 2,268 & 2,448 & 292 & 292 & 292 \\
\hyperlink{R53}{R53} & 1,527 & 1,519 & 1,538 & 292 & 292 & 292 \\
\hyperlink{R54}{R54} & 2,044 & 1,988 & 2,047 & 292 & 292 & 292 \\
\hyperlink{R58}{R58} & 2,790 & 1,779 & 2,871 & 292 & 292 & 292 \\
\hyperlink{R59}{R59} & 2,182 & 2,180 & 2,206 & 292 & 292 & 292 \\
\hyperlink{R60}{R60} & 2,306 & 1,603 & 2,352 & 292 & 292 & 292 \\
\hyperlink{R61}{R61} & 2,765 & 2,721 & 2,966 & 292 & 292 & 292 \\
\hyperlink{R62}{R62} & 2,453 & 2,438 & 2,493 & 292 & 292 & 292 \\
\hyperlink{R63}{R63} & 2,383 & 1,608 & 2,438 & 292 & 292 & 292 \\
\hyperlink{R64}{R64} & 1,514 & 1,504 & 1,527 & 292 & 292 & 292 \\
\hyperlink{R66}{R66} & 355 & 355 & 369 & 292 & 292 & 292 \\
\hyperlink{R67}{R67} & 358 & 290 & 380 & 292 & 292 & 292 \\
\hyperlink{R68}{R68} & 1,266 & 1,225 & 1,285 & 292 & 292 & 292 \\
\hyperlink{R69}{R69} & 1,677 & 1,659 & 1,686 & 292 & 292 & 292 \\
\hyperlink{R70}{R70} & 1,960 & 1,953 & 1,972 & 292 & 292 & 292 \\
\hyperlink{R71}{R71} & 1,682 & 1,617 & 1,825 & 292 & 292 & 292 \\
\hyperlink{R73}{R73} & 1,339 & 1,047 & 1,576 & 292 & 292 & 292 \\
\bottomrule\end{longtable}
\subsection{Kornia GPU}
\begin{longtable}{lrrrrrr}
\textbf{Recipe} & \multicolumn{3}{c}{\textbf{Throughput (images/s)}} & \multicolumn{3}{c}{\textbf{GPU memory (MiB)}} \\
 & Median & Min & Max & Median & Min & Max \\ \midrule \endhead
\hyperlink{R01}{R01} & 2,737 & 2,503 & 3,770 & 316 & 316 & 316 \\
\hyperlink{R02}{R02} & 2,393 & 2,356 & 2,473 & 316 & 316 & 316 \\
\hyperlink{R03}{R03} & 2,046 & 2,013 & 2,124 & 316 & 316 & 316 \\
\hyperlink{R04}{R04} & 2,396 & 2,393 & 2,504 & 390 & 390 & 390 \\
\hyperlink{R05}{R05} & 2,537 & 2,373 & 2,554 & 390 & 390 & 390 \\
\hyperlink{R06}{R06} & 1,968 & 1,896 & 2,003 & 316 & 316 & 316 \\
\hyperlink{R07}{R07} & 2,344 & 1,619 & 2,462 & 1,004 & 1,004 & 1,004 \\
\hyperlink{R08}{R08} & 2,497 & 2,371 & 2,514 & 1,004 & 1,004 & 1,004 \\
\hyperlink{R09}{R09} & 2,443 & 2,363 & 2,466 & 980 & 980 & 980 \\
\hyperlink{R10}{R10} & 2,325 & 1,671 & 2,451 & 1,154 & 1,154 & 1,154 \\
\hyperlink{R11}{R11} & 2,372 & 2,365 & 2,385 & 1,804 & 1,776 & 1,804 \\
\hyperlink{R12}{R12} & 2,417 & 2,383 & 2,435 & 390 & 390 & 390 \\
\hyperlink{R13}{R13} & 2,395 & 1,699 & 2,418 & 442 & 442 & 442 \\
\hyperlink{R14}{R14} & 2,378 & 1,576 & 2,448 & 464 & 464 & 464 \\
\hyperlink{R15}{R15} & 2,462 & 2,428 & 2,492 & 910 & 910 & 910 \\
\hyperlink{R16}{R16} & 2,408 & 2,346 & 2,510 & 760 & 760 & 760 \\
\hyperlink{R17}{R17} & 2,391 & 1,617 & 2,443 & 390 & 390 & 390 \\
\hyperlink{R18}{R18} & 2,374 & 2,344 & 2,404 & 724 & 724 & 724 \\
\hyperlink{R19}{R19} & 2,317 & 2,316 & 2,492 & 686 & 686 & 686 \\
\hyperlink{R20}{R20} & 2,395 & 2,293 & 2,410 & 946 & 946 & 946 \\
\hyperlink{R21}{R21} & 2,410 & 2,344 & 2,542 & 464 & 464 & 464 \\
\hyperlink{R22}{R22} & 2,520 & 2,401 & 2,626 & 1,022 & 1,022 & 1,022 \\
\hyperlink{R23}{R23} & 2,339 & 1,610 & 2,467 & 676 & 676 & 676 \\
\hyperlink{R24}{R24} & 2,431 & 2,297 & 2,470 & 1,824 & 1,824 & 1,824 \\
\hyperlink{R25}{R25} & 2,473 & 2,333 & 2,483 & 538 & 538 & 538 \\
\hyperlink{R26}{R26} & 2,446 & 1,607 & 2,495 & 590 & 590 & 590 \\
\hyperlink{R27}{R27} & 793 & 790 & 797 & 6,248 & 6,248 & 6,248 \\
\hyperlink{R28}{R28} & 2,400 & 2,386 & 2,484 & 468 & 468 & 468 \\
\hyperlink{R29}{R29} & 1,686 & 1,683 & 2,319 & 2,110 & 2,110 & 2,110 \\
\hyperlink{R30}{R30} & 2,525 & 2,381 & 2,549 & 464 & 464 & 464 \\
\hyperlink{R31}{R31} & 2,507 & 2,260 & 2,549 & 464 & 464 & 464 \\
\hyperlink{R32}{R32} & 2,472 & 1,681 & 2,497 & 540 & 540 & 540 \\
\hyperlink{R33}{R33} & 2,437 & 1,641 & 2,443 & 390 & 390 & 390 \\
\hyperlink{R34}{R34} & 2,391 & 2,348 & 2,465 & 1,156 & 1,156 & 1,156 \\
\hyperlink{R35}{R35} & 2,367 & 2,361 & 2,524 & 1,056 & 1,056 & 1,056 \\
\hyperlink{R36}{R36} & 2,435 & 2,432 & 2,574 & 958 & 958 & 958 \\
\hyperlink{R37}{R37} & 1,922 & 1,607 & 2,401 & 1,644 & 1,644 & 1,644 \\
\hyperlink{R38}{R38} & 2,399 & 2,022 & 2,552 & 2,604 & 2,604 & 2,604 \\
\hyperlink{R39}{R39} & 2,364 & 2,349 & 2,482 & 2,604 & 2,604 & 2,604 \\
\hyperlink{R40}{R40} & 2,447 & 1,694 & 2,477 & 1,160 & 1,160 & 1,160 \\
\hyperlink{R41}{R41} & 2,304 & 2,277 & 2,326 & 1,200 & 810 & 1,200 \\
\hyperlink{R42}{R42} & 800 & 787 & 814 & 614 & 614 & 614 \\
\hyperlink{R43}{R43} & 2,436 & 2,427 & 2,480 & 1,248 & 1,248 & 1,248 \\
\hyperlink{R44}{R44} & 2,447 & 2,430 & 2,474 & 1,150 & 1,150 & 1,150 \\
\hyperlink{R45}{R45} & 2,429 & 2,306 & 2,440 & 908 & 908 & 908 \\
\hyperlink{R46}{R46} & 2,458 & 1,678 & 2,474 & 930 & 930 & 930 \\
\hyperlink{R47}{R47} & 2,441 & 1,616 & 2,487 & 2,352 & 2,352 & 2,352 \\
\hyperlink{R49}{R49} & 2,350 & 2,350 & 2,373 & 1,792 & 1,776 & 1,866 \\
\hyperlink{R50}{R50} & 1,368 & 1,338 & 1,378 & 316 & 316 & 316 \\
\hyperlink{R51}{R51} & 993 & 789 & 1,017 & 316 & 316 & 316 \\
\hyperlink{R52}{R52} & 2,519 & 2,499 & 2,595 & 390 & 390 & 390 \\
\hyperlink{R53}{R53} & 2,503 & 2,375 & 2,562 & 1,004 & 1,004 & 1,004 \\
\hyperlink{R54}{R54} & 2,481 & 2,390 & 2,534 & 464 & 464 & 464 \\
\hyperlink{R58}{R58} & 1,934 & 1,839 & 2,935 & 316 & 316 & 316 \\
\hyperlink{R59}{R59} & 2,233 & 1,458 & 2,265 & 412 & 412 & 412 \\
\hyperlink{R60}{R60} & 2,419 & 2,419 & 2,443 & 390 & 390 & 390 \\
\hyperlink{R61}{R61} & 2,677 & 2,673 & 2,723 & 316 & 316 & 316 \\
\hyperlink{R62}{R62} & 2,482 & 2,416 & 2,625 & 316 & 316 & 316 \\
\hyperlink{R63}{R63} & 2,430 & 2,401 & 2,478 & 390 & 390 & 390 \\
\hyperlink{R64}{R64} & 2,403 & 2,360 & 2,510 & 714 & 714 & 714 \\
\hyperlink{R66}{R66} & 2,025 & 1,534 & 2,097 & 2,900 & 2,900 & 2,900 \\
\hyperlink{R67}{R67} & 2,042 & 2,029 & 2,085 & 2,900 & 2,900 & 2,900 \\
\hyperlink{R68}{R68} & 1,208 & 1,176 & 1,295 & 316 & 316 & 316 \\
\hyperlink{R69}{R69} & 2,208 & 1,538 & 2,277 & 390 & 390 & 390 \\
\hyperlink{R70}{R70} & 2,419 & 2,401 & 2,474 & 676 & 676 & 676 \\
\hyperlink{R71}{R71} & 2,393 & 2,385 & 2,425 & 466 & 466 & 466 \\
\hyperlink{R73}{R73} & 2,507 & 2,362 & 2,522 & 930 & 930 & 930 \\
\bottomrule\end{longtable}
\subsection{DALI GPU}
\begin{longtable}{lrrrrrr}
\textbf{Recipe} & \multicolumn{3}{c}{\textbf{Throughput (images/s)}} & \multicolumn{3}{c}{\textbf{GPU memory (MiB)}} \\
 & Median & Min & Max & Median & Min & Max \\ \midrule \endhead
\hyperlink{R01}{R01} & 5,078 & 5,055 & 5,089 & 1,808 & 1,808 & 1,808 \\
\hyperlink{R02}{R02} & 5,042 & 5,040 & 5,047 & 1,808 & 1,808 & 1,808 \\
\hyperlink{R03}{R03} & 5,087 & 5,072 & 5,099 & 1,808 & 1,808 & 1,808 \\
\hyperlink{R04}{R04} & 5,015 & 5,001 & 5,057 & 1,808 & 1,808 & 1,808 \\
\hyperlink{R05}{R05} & 5,055 & 4,990 & 5,067 & 1,808 & 1,808 & 1,808 \\
\hyperlink{R06}{R06} & 5,019 & 4,993 & 5,020 & 1,872 & 1,872 & 1,872 \\
\hyperlink{R07}{R07} & 5,033 & 4,989 & 5,035 & 1,872 & 1,872 & 1,872 \\
\hyperlink{R08}{R08} & 5,041 & 4,997 & 5,089 & 1,872 & 1,872 & 1,872 \\
\hyperlink{R11}{R11} & 5,037 & 4,995 & 5,042 & 1,872 & 1,872 & 1,872 \\
\hyperlink{R12}{R12} & 4,927 & 4,918 & 4,939 & 1,872 & 1,872 & 1,872 \\
\hyperlink{R13}{R13} & 5,046 & 5,043 & 5,052 & 1,872 & 1,872 & 1,872 \\
\hyperlink{R14}{R14} & 4,916 & 4,904 & 4,943 & 1,872 & 1,872 & 1,872 \\
\hyperlink{R15}{R15} & 4,999 & 4,987 & 5,003 & 1,872 & 1,872 & 1,872 \\
\hyperlink{R16}{R16} & 5,024 & 4,992 & 5,027 & 1,872 & 1,872 & 1,872 \\
\hyperlink{R17}{R17} & 5,047 & 5,025 & 5,057 & 1,872 & 1,872 & 1,872 \\
\hyperlink{R18}{R18} & 5,018 & 5,017 & 5,093 & 1,872 & 1,872 & 1,872 \\
\hyperlink{R19}{R19} & 5,054 & 4,984 & 5,082 & 1,872 & 1,872 & 1,872 \\
\hyperlink{R22}{R22} & 5,014 & 4,993 & 5,038 & 1,872 & 1,872 & 1,872 \\
\hyperlink{R23}{R23} & 4,981 & 4,969 & 5,030 & 1,872 & 1,872 & 1,872 \\
\hyperlink{R24}{R24} & 5,052 & 5,034 & 5,055 & 1,872 & 1,872 & 1,872 \\
\hyperlink{R25}{R25} & 5,037 & 4,972 & 5,060 & 1,872 & 1,872 & 1,872 \\
\hyperlink{R27}{R27} & 4,222 & 4,219 & 4,225 & 1,874 & 1,874 & 1,874 \\
\hyperlink{R29}{R29} & 4,875 & 4,870 & 4,882 & 1,872 & 1,872 & 1,872 \\
\hyperlink{R30}{R30} & 5,029 & 4,990 & 5,035 & 1,872 & 1,872 & 1,872 \\
\hyperlink{R31}{R31} & 5,035 & 5,019 & 5,046 & 1,872 & 1,872 & 1,872 \\
\hyperlink{R32}{R32} & 5,006 & 4,995 & 5,010 & 1,872 & 1,872 & 1,872 \\
\hyperlink{R33}{R33} & 5,021 & 5,005 & 5,023 & 1,872 & 1,872 & 1,872 \\
\hyperlink{R37}{R37} & 5,039 & 5,031 & 5,071 & 1,872 & 1,872 & 1,872 \\
\hyperlink{R42}{R42} & 5,000 & 4,987 & 5,023 & 1,872 & 1,872 & 1,872 \\
\hyperlink{R43}{R43} & 5,073 & 5,001 & 5,095 & 1,872 & 1,872 & 1,872 \\
\hyperlink{R46}{R46} & 5,032 & 5,019 & 5,056 & 1,872 & 1,872 & 1,872 \\
\hyperlink{R49}{R49} & 4,986 & 4,922 & 4,992 & 1,872 & 1,872 & 1,872 \\
\hyperlink{R50}{R50} & 5,025 & 4,965 & 5,037 & 1,872 & 1,872 & 1,872 \\
\hyperlink{R51}{R51} & 4,995 & 4,984 & 4,999 & 2,000 & 1,936 & 2,128 \\
\hyperlink{R52}{R52} & 5,021 & 5,014 & 5,022 & 2,128 & 2,128 & 2,128 \\
\hyperlink{R53}{R53} & 5,014 & 5,003 & 5,029 & 2,128 & 2,128 & 2,128 \\
\hyperlink{R58}{R58} & 5,033 & 5,029 & 5,044 & 2,128 & 2,128 & 2,128 \\
\hyperlink{R59}{R59} & 5,056 & 5,054 & 5,087 & 2,128 & 2,128 & 2,128 \\
\hyperlink{R60}{R60} & 5,092 & 5,051 & 5,100 & 2,128 & 2,128 & 2,128 \\
\hyperlink{R61}{R61} & 5,070 & 5,042 & 5,074 & 2,128 & 2,128 & 2,128 \\
\hyperlink{R62}{R62} & 4,992 & 4,988 & 5,005 & 2,128 & 2,128 & 2,128 \\
\hyperlink{R63}{R63} & 5,043 & 5,038 & 5,052 & 2,128 & 2,128 & 2,128 \\
\hyperlink{R64}{R64} & 5,060 & 5,048 & 5,089 & 2,128 & 2,128 & 2,128 \\
\hyperlink{R66}{R66} & 5,047 & 5,031 & 5,073 & 2,130 & 2,130 & 2,130 \\
\hyperlink{R67}{R67} & 5,064 & 5,049 & 5,076 & 2,130 & 2,130 & 2,130 \\
\hyperlink{R68}{R68} & 4,961 & 4,959 & 4,978 & 2,448 & 2,256 & 2,448 \\
\hyperlink{R69}{R69} & 5,035 & 4,995 & 5,036 & 2,448 & 2,448 & 2,448 \\
\hyperlink{R70}{R70} & 5,042 & 5,018 & 5,051 & 2,448 & 2,448 & 2,448 \\
\hyperlink{R71}{R71} & 5,042 & 5,029 & 5,055 & 2,448 & 2,448 & 2,448 \\
\hyperlink{R72}{R72} & 5,054 & 5,052 & 5,076 & 2,448 & 2,448 & 2,448 \\
\hyperlink{R73}{R73} & 5,078 & 5,074 & 5,083 & 2,448 & 2,448 & 2,448 \\
\bottomrule\end{longtable}
\subsection{Pillow-AllSIMD CPU (fast)}
\begin{longtable}{lrrrrrr}
\textbf{Recipe} & \multicolumn{3}{c}{\textbf{Throughput (images/s)}} & \multicolumn{3}{c}{\textbf{GPU memory (MiB)}} \\
 & Median & Min & Max & Median & Min & Max \\ \midrule \endhead
\hyperlink{R01}{R01} & 3,991 & 3,844 & 5,825 & 330 & 330 & 330 \\
\hyperlink{R02}{R02} & 4,437 & 4,374 & 4,544 & 330 & 330 & 330 \\
\hyperlink{R03}{R03} & 3,836 & 3,730 & 3,862 & 330 & 330 & 330 \\
\hyperlink{R04}{R04} & 4,187 & 2,731 & 4,418 & 330 & 330 & 330 \\
\hyperlink{R05}{R05} & 4,272 & 4,254 & 4,532 & 330 & 330 & 330 \\
\hyperlink{R06}{R06} & 3,990 & 3,840 & 4,021 & 330 & 330 & 330 \\
\hyperlink{R07}{R07} & 4,140 & 4,084 & 4,165 & 330 & 330 & 330 \\
\hyperlink{R08}{R08} & 3,995 & 2,441 & 4,016 & 330 & 330 & 330 \\
\hyperlink{R12}{R12} & 4,325 & 4,122 & 4,490 & 330 & 330 & 330 \\
\hyperlink{R13}{R13} & 2,703 & 2,586 & 4,343 & 330 & 330 & 330 \\
\hyperlink{R14}{R14} & 3,974 & 3,908 & 4,252 & 330 & 330 & 330 \\
\hyperlink{R15}{R15} & 3,750 & 3,668 & 3,807 & 330 & 330 & 330 \\
\hyperlink{R17}{R17} & 4,186 & 4,111 & 4,374 & 330 & 330 & 330 \\
\hyperlink{R18}{R18} & 4,153 & 4,114 & 4,213 & 330 & 330 & 330 \\
\hyperlink{R19}{R19} & 4,214 & 2,551 & 4,222 & 330 & 330 & 330 \\
\hyperlink{R21}{R21} & 3,883 & 3,840 & 3,913 & 330 & 330 & 330 \\
\hyperlink{R22}{R22} & 3,917 & 3,792 & 4,048 & 330 & 330 & 330 \\
\hyperlink{R24}{R24} & 3,560 & 3,539 & 3,600 & 330 & 330 & 330 \\
\hyperlink{R25}{R25} & 4,075 & 4,061 & 4,115 & 330 & 330 & 330 \\
\hyperlink{R27}{R27} & 3,920 & 3,812 & 4,014 & 330 & 330 & 330 \\
\hyperlink{R30}{R30} & 3,848 & 2,425 & 3,923 & 330 & 330 & 330 \\
\hyperlink{R31}{R31} & 3,707 & 2,365 & 3,859 & 330 & 330 & 330 \\
\hyperlink{R32}{R32} & 4,244 & 4,042 & 4,258 & 330 & 330 & 330 \\
\hyperlink{R33}{R33} & 4,265 & 4,204 & 4,537 & 330 & 330 & 330 \\
\hyperlink{R43}{R43} & 3,906 & 3,688 & 3,986 & 330 & 330 & 330 \\
\hyperlink{R46}{R46} & 3,713 & 2,472 & 3,811 & 330 & 330 & 330 \\
\hyperlink{R50}{R50} & 2,965 & 2,964 & 3,017 & 330 & 330 & 330 \\
\hyperlink{R51}{R51} & 2,388 & 2,341 & 2,404 & 330 & 330 & 330 \\
\hyperlink{R52}{R52} & 4,322 & 4,279 & 4,481 & 330 & 330 & 330 \\
\hyperlink{R53}{R53} & 4,165 & 2,727 & 4,392 & 330 & 330 & 330 \\
\hyperlink{R55}{R55} & 4,068 & 3,867 & 4,152 & 330 & 330 & 330 \\
\hyperlink{R56}{R56} & 3,998 & 2,521 & 4,139 & 330 & 330 & 330 \\
\hyperlink{R57}{R57} & 3,511 & 3,503 & 3,561 & 330 & 330 & 330 \\
\hyperlink{R58}{R58} & 4,395 & 4,316 & 4,507 & 330 & 330 & 330 \\
\hyperlink{R59}{R59} & 3,814 & 3,691 & 3,920 & 330 & 330 & 330 \\
\hyperlink{R60}{R60} & 3,949 & 3,765 & 4,118 & 330 & 330 & 330 \\
\hyperlink{R61}{R61} & 3,910 & 3,766 & 3,989 & 330 & 330 & 330 \\
\hyperlink{R62}{R62} & 4,439 & 4,060 & 4,570 & 330 & 330 & 330 \\
\hyperlink{R63}{R63} & 3,910 & 3,905 & 3,936 & 330 & 330 & 330 \\
\hyperlink{R64}{R64} & 3,576 & 3,410 & 3,610 & 330 & 330 & 330 \\
\hyperlink{R65}{R65} & 3,591 & 3,549 & 3,667 & 330 & 330 & 330 \\
\hyperlink{R68}{R68} & 2,792 & 2,723 & 2,805 & 330 & 330 & 330 \\
\hyperlink{R69}{R69} & 3,608 & 3,598 & 3,623 & 330 & 330 & 330 \\
\hyperlink{R70}{R70} & 3,838 & 2,779 & 4,014 & 330 & 330 & 330 \\
\hyperlink{R71}{R71} & 3,731 & 3,713 & 3,784 & 330 & 330 & 330 \\
\hyperlink{R73}{R73} & 3,506 & 3,457 & 3,541 & 330 & 330 & 330 \\
\bottomrule\end{longtable}

\endgroup

\clearpage
\section{Transformation coverage}
\label{app:coverage}

The November 2026 census starts from \CoverageAlbumentationsX{} selected
2D RGB entries in AX. Its release versions are reported separately from
the benchmark runtime versions. Counts describe counterparts to this
catalog, including entries without a timed recipe.

\begin{table}[H]
\centering
\rowcolors{2}{gray!5}{white}
\begin{tabular}{lrr}
\textbf{Library} & \textbf{Matched AX catalog entries} & \textbf{Census version} \\ \midrule
\inputrows generated/coverage.tex
\bottomrule
\end{tabular}
\caption{Coverage relative to the AX catalog, checked against the listed
releases. Each count gives the number of AX entries with a counterpart.}
\label{tab:coverage}
\end{table}

The mapping includes public transforms, functions, and combinations of
named operations with a similar image effect. Sampling, border handling,
and annotation support can differ. One API can cover several entries:
TorchVision ColorJitter is listed for both ColorJitter and
RandomBrightnessContrast.

Each row is an AX catalog entry. A dash means the catalog records no
corresponding API for that library.

\begingroup\small
\rowcolors{2}{gray!5}{white}
\begin{longtable}{P{4.1cm}P{4.0cm}P{3.8cm}P{3.4cm}}
\textbf{AX entry} & \textbf{Kornia} & \textbf{TorchVision} & \textbf{Pillow} \\ \midrule\endhead
\inputrows generated/coverage-mapping.tex
\bottomrule
\end{longtable}
\endgroup

\clearpage
\section{Reproducing the reported measurements}
\label{app:reproduce}

The \repo{} contains the report generators and the immutable manifests.
The generators check each cell against its manifest and recipe checksum
and validate that the report matrix is complete.

\begin{table}[H]
\centering\small
\rowcolors{2}{gray!5}{white}
\begin{tabular}{ll}
\textbf{Implementation} & \textbf{Version recorded in cells} \\ \midrule
\inputrows generated/versions.tex
\bottomrule
\end{tabular}
\caption{Implementation versions recorded in the cells. The locks pin
PyTorch 2.15.0, Triton 3.9.0, SimpleJPEG 1.9.0, and the OpenCV Python wheel
5.0.0.93.}
\end{table}

The benchmark used Python 3.13.14 on
\nolinkurl{common-cu129-ubuntu-2204-nvidia-580-v20260730}. The cells are under
\nolinkurl{gs://imagenet_validation/augmentation-benchmark/runs/}, in the
following run directories:

\begin{description}
\item[Full RGB] \texttt{\expandafter\seqsplit\expandafter{\FullRunID}}\\
  Source \texttt{\FullSourceRevision}.
\item[AX 2.6.3] \texttt{\expandafter\seqsplit\expandafter{\AXRunID}}\\
  Source \texttt{\AXSourceRevision}.
\end{description}

The repository's \texttt{paper/README.md} gives the commands to regenerate the
report from the source archives and cells. The CSV in
\texttt{paper/generated/recipe-results.csv} records the source run ID, cell IDs,
versions, declared stages, throughput, and GPU memory for every seed.
The separate coverage catalog has SHA-256
\texttt{\expandafter\seqsplit\expandafter{\CoverageCatalogSHA}}.

\end{document}